\documentclass[pra, reprint, showpacs, showkeys, nofootinbib]{revtex4-1}
\usepackage{graphicx}
\usepackage[colorlinks=true, citecolor=blue, urlcolor=blue ]{hyperref}
\usepackage{amsmath, amssymb}

\newcommand{\bq}{\begin{equation}}
\newcommand{\eq}{\end{equation}}
\newcommand{\ba}{\begin{array}}
\newcommand{\ea}{\end{array}}
\newcommand{\bqa}{\begin{eqnarray}}
\newcommand{\eqa}{\end{eqnarray}}

\newcommand{\ran}{\rangle}
\newcommand{\lan}{\langle}
\newtheorem{theorem}{Theorem}

\begin{document}
\title{Optimal reducibility of all Stochastic Local Operation and Classical Communication equivalent $W$ states}
\author{Swapan Rana}
\email{swapanqic@gmail.com}
\author{Preeti Parashar}
\email{parashar@isical.ac.in}

\affiliation{Physics and Applied Mathematics Unit, Indian Statistical Institute, 203 BT Road, Kolkata, India}
\begin{abstract}

We show that all multipartite pure states that are SLOCC equivalent to the $N$-qubit $W$ state,
can be uniquely determined (among arbitrary states) from their bipartite marginals. We also prove that
only $(N-1)$ of the bipartite marginals are sufficient and this is also the optimal number.
Thus, contrary to the $GHZ$ class, $W$-type states preserve their reducibility under SLOCC. We also study the optimal reducibility of some larger classes of states. The generic Dicke states $|GD_N^\ell\ran$ are shown to be optimally determined by their ($\ell+1$)-partite marginals. The class of `$G$' states (superposition of $W$ and $\bar{W}$) are shown to be optimally determined by just two $(N-2)$-partite marginals.
\end{abstract}
\pacs{03.67.Mn, 03.65.Ud }
\keywords{SLOCC,  reduced density matrices, (ir)reducible correlations}
\date{\today}
\maketitle

\section{introduction}
Characterization of multipartite entanglement is an interesting field of study in quantum information. Although for bipartite states  (particularly for
pure states), nearly all aspects of entanglement have been understood, there still remain lots of unresolved issues in characterization, manipulation and
quantification of multipartite entanglement. For multipartite state, not only the amount but also the \emph{flavor} of entanglement becomes a complicated issue and there are various perspectives to study entanglement at the multiparty level, such as, its characterization
by means of local operations and classical communications (LOCC), its ability to reject local realism and hidden variable theories, etc. It is interesting to find any possible relationship between different perspectives or at least to know how the well known states behaves in different perspectives. In this article, we will try to explore some relations between two different perspectives.

The central issue in LOCC characterization is the convertibility between different multipartite states using LOCC. If the states can be converted to each other with a non-zero probability, the two states are called stochastic LOCC (SLOCC) equivalent and they represent the same flavor of entanglement. For example, in case of three qubits, there are only two \emph{kinds} of inequivalent genuine tripartite entanglement represented by the Greenberger-Horne-Zeilinger (GHZ) and $W$-type entanglement \cite{dvc2000}. The present article concerns mainly with all multiqubit $W$-type states.

From another well studied perspective, namely,`\emph{Parts and Whole}', the basic question is that of \emph{reducibility} of the correlation exhibited by composite quantum states \cite{lpw2002,lw2002,jl2005,wl2009,z2008,pr2009,prftc2009}.
In this qualitative approach ( \cite{z2008} deals with a parallel quantitative approach) the flavor of entanglement in a composite state depends
on the \emph{determinability} of the state by its subsystems. Precisely, if a state can be determined uniquely (among arbitrary states) by a set of its $K$-partite reduced density matrices (RDMs) but not by $(K-1)$-partite RDMs, the correlation in the state is said to be \emph{reducible} to $K$-partite level. Jones and Linden \cite{jl2005} have shown that the correlation in \emph{almost all} $N$-qudit pure states
 is reducible to $(\lceil N/2\rceil+1)$-partite level .

Recently it has been shown by Walck and Lyons \cite{wl2009} that any $N$-qubit pure state is not determined by its RDMs if and only if it is local unitary (LU) equivalent to the \emph{generalized} $GHZ$ state \bq\label{int1}|GGHZ\ran=a|00\ldots0\ran+b|11\ldots1\ran.\eq Obviously, $|GGHZ\ran$ is not necessarily LU for all $a, b$ but is always SLOCC equivalent to the \emph{standard} $GHZ$ state (the one with $a=b=1/\sqrt{N}$)\footnote{For the simplest case of $N=2$, it follows from Nielsen's majorization criterion \cite{nielsen99} that $|GGHZ\ran$ can not be converted to $|GHZ\ran$ even by LOCC.}. Therefore, any state which is SLOCC equivalent to the $GHZ$ state should be SLOCC equivalent to (\ref{int1}). Very recently Ref.\cite{dsrar} considers the question whether each SLOCC equivalent $GHZ$ state preserves the irreducibility property of $|GGHZ\ran$. Since LU $\subset$ SLOCC, there always exist pure states which are SLOCC but not LU equivalent to $|GGHZ\ran$ and it follows from \cite{wl2009} that such states are determined by their RDMs. Thus $GHZ$-type entangled states can not preserve their (ir)reducibility under SLOCC.

 The other well known class of pure states which has been extensively studied both theoretically as well as experimentally is the $W$ class. We have recently shown \cite{pr2009} that the  $W$-class of states are completely determined by just their two-party RDMs.
So, in view of the above, a natural query arises about the reducibility of all those states which are SLOCC equivalent to
$W$ state. Moreover, what is the minimum number of RDMs required to determine such states?  As the $GHZ$ example shows, there is no guarantee that the reducibility of a state from its RDMs is preserved under SLOCC. So, the reducible correlations in all SLOCC equivalent $W$ states is worth exploring. These questions have motivated the present article and surprisingly we find that the SLOCC equivalent $W$-type states are also determined by just their bipartite RDMs, thereby preserving reducibility. This is yet another peculiar property of $W$-type states not exhibited by $GHZ$-type states.

Another motivation for investigating SLOCC equivalent states stems from the attention they have received in recent literature. To mention a few, a widely used entanglement measure, the geometric measure of entanglement has been generalized to distances from the set of product states  to the set that remains invariant under SLOCC \cite{ut10}. In the study of entanglement manipulation, the notion of entanglement assisted multi-copy LOCC transformation (eMLOCC) has recently been extended to its stochastic version (eMSLOCC) \cite{winter10}. Here, for the sake of generality, we will consider the \emph{generic} $W$-type states (instead of the \emph{standard} one having all coefficients $1/\sqrt{N}$). It is known that some of these states can be used for perfect teleportation and superdense coding while the standard $W$ state cannot \cite{exp}.

The organization of this article is as follows. In Sec. \ref{w}, we briefly describe a canonical form of $W$-type states.
The main results for such states are described in Sec. \ref{main}. In Sec. \ref{iv}, we generalize the result of $W$ states to some other classes of states. To be precise, the optimal reducibility of the \emph{generic} Dicke states $|D_N^\ell\ran$ and $|G_N\ran$ state (superposition of $W$ and $\bar{W}$) is investigated. We conclude in Sec. \ref{end} after a discussion on possible extensions of the results obtained.
\section{A canonical form of W-type States}\label{w}
In \cite{pr2009}, we have considered the following class of states as `generalized $W$ states':
\bq\label{gw1}|W\ran=w_1|10\ldots0\ran+w_2|01\ldots0\ran+\ldots+w_N|00\ldots1\ran\eq Clearly the state in (\ref{gw1})
is of `$W$-type' as the SLOCC operator $\otimes_{k=1}^{N}A_k$ with $$A_k=|0\ran\lan0|+\frac{1}{\sqrt{N}w_k}|1\ran\lan1|$$
 transforms it into the \textit{standard} $N$-qubit $W$ state \bq\label{nw1}|W_N\ran=\frac{1}{\sqrt{N}}
(|10\ldots0\ran+|01\ldots0\ran+\ldots+|00\ldots1\ran).\eq

However, the purpose of the present article is to consider all possible $W$-type states. So we want a convenient canonical form
for all such states. To derive the desired form, we shall follow the treatment of Ref. \cite{kt2010}. Any $N$-qubit pure state which is
SLOCC equivalent to the standard $W$ state (\ref{nw1}) is given by \bq\label{slw1}|\psi\ran=\otimes_{k=1}^{N}A_k|W_N\ran\eq
where $A_i$s are any invertible operators. If $$A_k=\left[
                                                     \begin{array}{cc}
                                                       \alpha_k & \gamma_k \\
                                                       \beta_k & \delta_k \\
                                                     \end{array}
                                                   \right]$$
then $A_k$ transforms $|0\ran_k$ to $\alpha_k|0\ran_k+\beta_k|1\ran_k\equiv|u\ran_k$ and
$|1\ran_k$ to $\gamma_k|0\ran_k+\delta_k|1\ran_k\equiv|v\ran_k$ and so from (\ref{slw1})
\bq\label{slw2}|\psi\ran=\frac{1}{\sqrt{N}}(|vu\ldots u\ran+|uv\ldots u\ran+\ldots+|uu\ldots v\ran).\eq

Now the invertibility of $A_k$ implies $u_k$ and $v_k$ are independent and hence can be extended
to form an orthonormal basis of the local Hilbert space. Thus setting $p_k$ parallel to $u_k$ and $q_k$ orthonormal to $p_k$ by
\begin{eqnarray*}
|p\ran_k&=&a_k|u\ran_k\\
|q\ran_k&=&b_k|u\ran_k+b'_k|v\ran_k,
\end{eqnarray*}
(\ref{slw2}) becomes \bq\label{slw3} |\psi\ran=z_0|pp\ldots p\ran+\sum_{k=1}^Nz_k|pp\ldots p_{k-1}q_kp_{k+1}\ldots p\ran\eq

Clearly the bases can be redefined  to \textit{absorb} the phases in the complex coefficients and (\ref{slw3}) can be written as
\bq\label{slw4} |GW\ran=c_0|0\ldots 0\ran+\sum_{k=1}^Nc_k|0\ldots 0_{k-1}1_k0_{k+1}\ldots 0\ran,\eq
$c_k\ge0,\quad \sum_{k=0}^Nc_k^2=1$ (for normalization).

Thus any SLOCC equivalent $W$ state can be expressed as (\ref{slw4}) for some local orthonormal basis $\{|0\ran_k, |1\ran_k\}$ and
$c_i\ge0$. A detailed discussion on manipulation of this class of states under local operations has been carried out in \cite{kt2010}.

We note that if only one $c_k$ is non-zero, then, being a product state, it is uniquely determined  by its (single-partite) subsystems. Similarly if only two of the $c_k$s are non-zero, it is at most a bipartite entangled state and hence determined by its one bipartite (of the entangled parties) and all other single-partite (all the rest are in a product state) marginals. So, for non-trivial case, we can assume at least three of the $c_k$s to be non-zero and without loss of generality, let us assume $c_1\ne0$. Also, we will not restrict the coefficients to be real (though it will yield the same result), rather we will take them as \emph{arbitrary complex numbers}, satisfying normalization condition. The sought result for this class is stated below.
\section{Determination of $|GW\ran$ from bipartite marginals}\label{main}
\subsection{The main Result}
\begin{theorem}\label{th1}
All SLOCC equivalent W states are uniquely determined among arbitrary states
by their bipartite marginals $\rho^{1K}_{GW},\quad K=2(1)N$.
\end{theorem}

Here and henceforth the superscripts in RDMs indicate the constituent subsystems and the subscripts indicate the original state from which it has been calculated (e.g., here $|GW\ran$). Also the notation $K=2(1)N$ means $K$ is in the range 2 to $N$ with increments of 1 i.e., $K=2,3,\dotsc, N$. To prove the Theorem we will show that there does not exists any other (except the original $|GW\ran\lan GW|$) density
matrix sharing the same RDMs $\rho_{GW}^{1K},\quad K=2(1)N$.

 The main mathematical ingredients in the proof are some well-known properties \cite{hj1985} of positive semi definite (PSD) matrices : If a hermitian matrix $A=(a_{ij})$ be PSD (written as $A\ge0$), then \begin{itemize}
                                    \item[(i)] $a_{ii}\ge0\quad\forall i$.
                                    \item[(ii)] If some $a_{kk}=0$, then $a_{ik}=a_{kj}=0\quad\forall i,j$.
                                    \item[(iii)] $a_{ii}a_{jj}\ge |a_{ij}|^2\quad\forall i,j$.
                                    \item[(iv)] All \emph{principle minors}\footnote{Let $A$ be an $n\times n$ matrix and $S$ be a subset of the set $\{1,2,\ldots n\}$. Then the determinant of the submatrix obtained by deleting all the rows and columns of $A$ whose index are not in $S$, is called a principle minor of $A$. The principle minor consisting of the rows (and columns) $i_1,i_2,\ldots i_r$ is usually denoted by $A[i_1,i_2,\ldots i_r]$.}
                                    of $A$ are non-negative (this condition together with det$A\ge0$ is also a sufficient condition for PSD).
                                 \end{itemize}

\textbf{Proof:}

1.\quad From (\ref{slw4}), we readily have
\bq\label{pr1} \rho_{GW}^{1K}=\left[
                           \begin{array}{cccc}
                             n_{1K} & c_0\bar{c}_K & c_0\bar{c}_1 & 0 \\
                              & |c_K|^2 & c_K\bar{c}_1 & 0 \\
                              &  & |c_1|^2& 0 \\
                              &  &  & 0 \\
                           \end{array}
                         \right]
\eq
where $n_{1K}=1-|c_1|^2-|c_K|^2$ by normalization and we are showing only the upper-half
entries (as the upper-half $a_{ij}\quad\forall i\le j$ is a sufficient description of a hermitian matrix $A=a_{ij}$).

2.\quad Now, if possible, let another $N$-qubit density matrix (possibly mixed, thereby subscript $M$)\footnote{Instead of using  single index $i$ and $j$ to denote rows and columns of a matrix, a ``lexicographically ordered" multi index $(i_1i_2\ldots i_N)$ and $(j_1j_2\ldots j_N)$ has been used here, e.g., $r_{IJ}:=r_{(i_1i_2\ldots i_N)(j_1j_2\ldots j_N)}$.}
\bq\label{grp1}\rho_M=\sum_{i_1,\ldots,j_N=0}^{1}r_{(i_1\ldots i_N)(j_1\ldots j_N)}|i_1\ldots i_N\ran\lan j_1\ldots j_N|\eq
share the same bipartite RDMs with $|GW\ran$ i.e., $\rho_{M}^{1K}=\rho_{GW}^{1K}\quad\forall K=2(1)N$. For (\ref{grp1}) to represent a valid physical state, we must have $\bar{r}_{(i_1\ldots i_N)(j_1\ldots j_N)}=r_{(j_1\ldots j_N)(i_1\ldots i_N)}$ (for hermiticity)and $\sum_{i_1,\ldots,i_N=0}^1
r_{(i_1\ldots i_N)(i_1\ldots i_N)}=1$ (from normalization Tr$(\rho_M)=1$). In addition, all the above four properties (i)--(iv)  of PSD matrices must hold.

3(a).\quad From (\ref{pr1}), since there exists no term $|11\ran\lan11|$ in $\rho_{GW}^{1K}$, we must have
$r_{(1i_2i_3\ldots 1_k\ldots i_N)(1i_2i_3\ldots 1_k\ldots i_N)}=0$ and hence by property (ii) of PSD matrices, we have
\bq\label{pr2} r_{(1i_2\ldots 1_k\ldots i_N)(j_1j_2\ldots j_N)}=r_{(i_1i_2\ldots i_N)(1j_2\ldots 1_k\ldots j_N)}=0,\eq
for all $i_1,i_2,\ldots,i_N,j_1,j_2,\ldots,j_N=0,1$.

(b).\quad Comparing the coefficient of $|10\ran\lan 10|$ from $\rho_{M}^{1N}$ and $\rho_{GW}^{1N}$, it follows that
\bq\label{pr3} r_{(10\ldots 0)(10\ldots 0)}=|c_1|^2.\eq

4(a).\quad Now consider the non-diagonal element $|01\ran\lan10|$ of $\rho_{M}^{1K}$ and $\rho_{GW}^{1K},\quad K=2(1)N$. By virtue of (\ref{pr2}), we have
\bq\label{pr4} r_{(0\ldots0_{K-1}1_K0_{K+1}\ldots0)(100\ldots0)}=c_K\bar{c}_1\eq and hence by the property (iii) of PSD matrices with $i=(0\ldots0_{K-1}1_K0_{K+1}\ldots0)$ and $j=(100\ldots0)$ we have
\bq\label{pr5} r_{(0\ldots0_{K-1}1_K0_{K+1}\ldots0)(0\ldots0_{K-1}1_K0_{K+1}\ldots0)}\ge |c_K|^2.\eq

(b).\quad Similarly, comparing the coefficients of $|00\ran\lan10|$ and using (\ref{pr2}), it follows that
\bq\label{pr6} r_{(00\ldots0)(00\ldots0)}\ge |c_0|^2.\eq

(c).\quad Now from normalization ($\sum_{k=0}^N|c_k|^2=\sum_{i_1,i_2,\ldots,i_N=0}^1r_{(i_1i_2\ldots i_N)(i_1i_2\ldots i_N)}=1)$ and the property (i) of PSD matrices it follows that all the inequalities in (\ref{pr5}) and (\ref{pr6}) will be equalities; and each  $r_{(i_1i_2\ldots i_N)(i_1i_2\ldots i_N)}$ in which two or more $i_k$s are 1, is zero. So, by property (ii), $r_{(i_1i_2\ldots i_N)(j_1j_2\ldots j_N)}=r_{(j_1j_2\ldots j_N)(i_1i_2\ldots i_N)}=0$ whenever two or more
$i_k$s are 1.

(d).\quad Comparing the coefficients of $|00\ran\lan01|$ from $\rho_{M}^{1K}$ and $\rho_{GW}^{1K}$, we have
\bq\label{pr7} r_{(00\ldots0)(00\ldots01_K0\ldots0)}=c_0\bar{c}_K,\quad\forall K=2(1)N\eq

(e).\quad Thus, collecting all the results it follows that $\rho_M$ has the same form as $|GW\ran\lan GW|$ and they share the same diagonal elements, same elements along the row and column $(00\ldots0)$ and $(10\ldots0)$. The only remaining task is to prove \bq\label{pr8}r_{(0\ldots01_J0\ldots0)(0\ldots01_K0\ldots0)}=c_J\bar{c}_K\eq for $J>K=1(1)(N-1)$. This part is quite difficult, because no further condition can arise from sharing of the RDMs.

5.\quad  If $c_J\bar{c}_K=0$ (which means $r_{(0\ldots1_J\ldots0)(0\ldots1_J\ldots0)}.r_{(0\ldots1_K\ldots0)(0\ldots1_K\ldots0)}=0$) then by property (ii), (\ref{pr8}) follows trivially. Hence let us assume $c_J\bar{c}_K\ne0$. To complete the proof we will now apply property (iv) to $\rho_M$. Let us consider the following principle minor consisting of the rows and columns $(0\ldots01_J0\ldots0), (0\ldots01_K0\ldots0), (10\ldots0)$:
\bq\label{pr9} \left|\begin{array}{ccc}
                       |c_J|^2 & r & c_J\bar{c}_1 \\
                       \bar{r} & |c_K|^2 & c_K\bar{c}_1 \\
                       \bar{c}_Jc_1 & \bar{c}_Kc_1 & |c_1|^2
                     \end{array}\right|\eq
where $r=r_{(0\ldots01_J0\ldots0)(0\ldots01_K0\ldots0)}$. The value of this determinant is\footnote{To evaluate easily, divide  first row by $c_J$, first column by $\bar{c}_J$, second row by $c_K$, second column by $\bar{c}_K$,  third row by $c_1$ and third column by $\bar{c}_1$.} $-|c_J|^2|c_K|^2|c_1|^2|1-\frac{r}{c_J\bar{c}_K}|^2$. Since this should be non-negative, we have $r=c_J\bar{c}_K$.\hfill $\blacksquare$

We have adopted the algorithmic style of writing the above proof from our previous work  \cite{pr2009}, for better clarity. As a result, the proofs may look similar, however we emphasize that the present proof is essentially very much different for the following reasons: (i) The class of states considered in \cite{pr2009} has been extended here to its most generalized form, encompassing \emph{all} SLOCC equivalent states. Previously  it was assumed that $c_0=0$ and all other $c_k\ne0$. Here all $c_k\ge0$, thereby some $c_k$ may vanish. So the present class really consists of various subclasses.  (ii) In \cite{pr2009}, the knowledge of all ($^NC_2=N(N-1)/2$ in number) bipartite RDMs were used, which ensured that the coefficient of $|11\ran\lan11|$ in every $\rho^{JK}$ should vanish. However, for the sake of optimality here we are restricting to only $J=1$. So, we can not compare RDMs having $J\ne1$ and hence by the previous technique even the diagonals can not be determined. Thus the present technique is different from the previous one (indeed, it supersedes the previous technique). (iii) Lastly, it is worth mentioning that step 5 in the present proof (for determining non-diagonal elements) can be viewed as a \emph{matrix completion problem}--a well studied problem in Mathematics community. We have found that the solution to such kind of \emph{PSD completion} is unique. This new technique will be applied to other classes of states. \emph{We emphasize that without this PSD completion step, it is impossible to prove the results, as there will be no more constraints from sharing of the RDMs}.

Our result has a notable similarity with \emph{entanglement combing} \cite{yangeisert09} in which any multipartite pure state can be transformed into bipartite pairwise entangled states in a ``\emph{lossless fashion}", keeping one party common to all. Coincidently, the correlations in  $W$-type states are distributed into its parts in a similar way i.e., bipartitely. So the correlations therein can be thought of as automatically \emph{combed}.

\subsection{Optimal number of RDMs} In \cite{pr2009}, we have shown that the class of states (\ref{gw1}) is uniquely determined, among pure states, by only $(N-1)$ bipartite marginals and we asked whether it is the optimal number of bipartite RDMs needed to determine it. It follows from the present Theorem  that $(N-1)$ is indeed the optimal number, generically no such state can be determined from fewer RDMs (provided the state is a truly entangled state, which is guaranteed by the restriction $c_i\ne0~\forall i=1(1)N$) . As an example, for $N=4$, the class of states \bq\label{ex1} |W_4\ran=\sum_{k=1}^4w_i|0\ldots1_k\ldots0\ran,\quad w_i\in\mathbb{ C}-\{0\}\eq can not be uniquely determined by any set of 2 bipartite marginals \cite{pr2009}. This optimal requirement is certainly drastically less compared to the known general bound of $\lfloor N/2\rfloor$ number of $(\lceil N/2\rceil+1)$-partite marginals \cite{jl2005} (since each of the latter RDMs contain higher order correlations not captured by bipartite RDMs).

\begin{figure}
  \includegraphics[width=8.5cm]{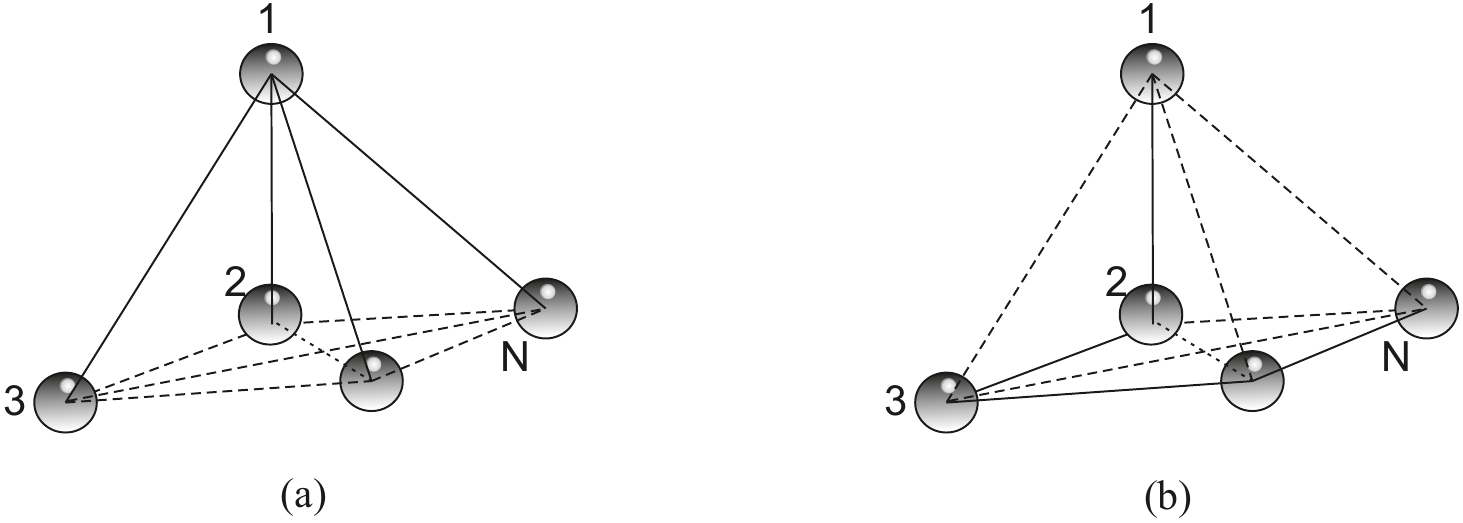}
  \caption{All $N$-qubit pure states which are SLOCC equivalent to $W$ states ($|GW\ran$ in the text) can be determined, among arbitrary states, by the following set of bipartite marginals: (a) $\{\rho^{1K}, K=2(1)N\}$; (b) $\{\rho^{K(K+1)},K=1(1)(N-1)\}$. Generally, in $|GW\ran$ each pair of qubits is correlated (e.g., entangled), which is indicated by the edges. The set of biparite RDMs required in each case is depicted by the solid edges. We note that the figure in (a) resembles \emph{entanglement combing} \cite{yangeisert09}.}\label{fig1}
\end{figure}

Though ($N$-1) is the optimal number, it is worth mentioning that there may exist other set of RDMs that can also determine these states\footnote{Unfortunately, we have not yet been able to characterize all states which are determined by $K$-partite RDMs. If this question can be answered, then all relevant queries can be settled. The $|GD_N^\ell\ran$ states are example of such states, for $2\le K\le\lfloor N/2 \rfloor$. It follows from \cite{wl2009} that except $|GGHZ\ran$ and its LU, all pure $N$-qubit states have $K\le N-1$.}. In the first attempt, a possible alternative set of RDMs is $\{\rho^{12}, \rho^{23}, \rho^{34},\ldots,\rho^{N-1~N}\}$. This is the argument of our next theorem. It is very likely that other similar sets are also sufficient. Since these sets of RDMs (necessarily covering all the parties) is the least possible RDMs from which a multipartite entangled state can be determined, we can say that the correlations in $|GW\ran$ can be reduced to lowest possible level (i.e., to bipartite order). This feature of $|W\ran$-type states is depicted in Fig. \ref{fig1}.

\begin{theorem}\label{th2}
All SLOCC equivalent W states are uniquely determined, among arbitrary states,
by their bipartite marginals $\rho^{K(K+1)}_{GW},\quad K=1(1)(N-1)$.
\end{theorem}
This is really surprising, because in this case the number of constraints are least possible (than all other previous cases). For example, the constraints for diagonal elements in $\rho^{J(J+1)}$ and $\rho^{(J+1)(J+2)}$ are almost redundant, as the coefficient of $|01\ran\lan01|$ in the first is exactly same to the coefficient of $|10\ran\lan10|$ in the later. This makes the proof very complicated and so we relegate it to the Appendix.

\section{Optimal reducibility of some other classes of states}\label{iv}
In this section, we will generalize the result on $W$ states to two other well known classes of states namely the Dicke states $|D_N^\ell\ran$ and the `$G$' state.

\subsection{Determination of $|D_N^{\ell}\ran$ from $(\ell+1)$-partite marginals}
 The \emph{generic} Dicke states are defined by \bq\label{defdicke} |GD_N^{\ell}\ran=\sum_i c_i|i\ran,\eq where $i=i_1i_2\ldots i_N$ and the sum varies over all permutations of $\ell (\le\lfloor N/2\rfloor)$ number of 1 and $N-\ell$ number of 0;  $ c_i\in \mathbb{C}$ with $\sum_i |c_i|^2=1$ and $c_i\ne0$. The \emph{standard} Dicke states (i.e., when all the coefficients are equal) are SLOCC inequivalent to each other for different $\ell$ and also inequivalent to the GHZ states \cite{dliepl09}. In our earlier work we had shown that for $\ell<\lfloor N/2\rfloor$, the class of states (\ref{defdicke}) is determined by $2\ell$-partite marginals. Also the question about its optimality was raised therein \cite{prftc2009}. It was discussed that the optimal number would lie between $[\ell+1,2\ell]$. In the spirit of our main result of the previous section, we hope that $\ell+1$ may be the optimal number. It indeed turns out to be the case. Thus we have the following general result:

\begin{theorem}\label{th3} The class of states given by (\ref{defdicke}) is uniquely determined, among arbitrary states, by its $(\ell+1)$-partite marginals $\rho^{1P_2P_3\ldots P_{\ell+1}}_{GD},\quad P_k\in \{2,3,4,\ldots,N\}$.\end{theorem}

\textbf{Proof:} The proof follows almost parallel to the case of $W$ states. We have to use PSD completion (third order minor) to show the uniqueness of the off-diagonals.\hfill $\blacksquare$

Note that we have used all the ($\ell+1$)-partite RDMs ($^{N-1}C_\ell$ in number). Surely not all of them are required and some are redundant. But we don't know what is the optimal number of ($\ell+1$)-partite RDMs. It follows trivially that this number can not be less than $N-\ell$.

\subsection{Optimal reducibility of the state $|G_N\ran=\frac{1}{\sqrt{2}}(|W_N\ran+|\bar{W}_N\ran)$}
Recently the correlation structure in the $N$-qubit `$G$' state, \bq\label{etan}|G_N\ran=\frac{1}{\sqrt{2}}(|W_N\ran+|\bar{W}_N\ran)\eq has been studied by several authors from different perspectives (for example see \cite{ausen03, ausen031, bennett11}). The purpose of this subsection is to consider the correlation structure in $|G_N\ran$ from the parts and whole point of view i.e., to determine its reducibility. This question has been raised and partially answered in \cite{ausen031}. The authors showed that for $N\ge5$, $|G_N\ran$ can be determined from $(N-1)$-partite RDMs. Here we show that the correlation is further reducible. For the sake of optimality we prove that for $N\ge6$, the correlation in $|G_N\ran$ is reducible to $(N-2)$-partite level and not beyond it.

The reducibility of three qubit $G$ state has been considered in  \cite{dsrar}. In fact this case follows trivially from the previous known results. It has been explicitly shown in \cite{lpw2002} that except the $|GGHZ\ran$ and its LU equivalent states, all three-qubit pure states are determined by their bipartite RDMs. So any pure state which is SLOCC equivalent to $W$ state can be determined as it is not LU equivalent to $|GGHZ\ran$. All other genuinely entangled states which are SLOCC but not LU equivalent to $|GGHZ\ran$ are also determinable. Such states provide examples to a query raised in \cite{pr2009}. As an instant example, $|G_3\ran$ is such a state as it is not LU\footnote{Partial transpose of any (the state is symmetric) bipartite RDM of $|G_3\ran$ has a negative eigenvalue -1/6, so (by PPT criterion) is entangled. However, any bipartite RDM of $|GGHZ\ran$ is separable. Therefore they cannot be LU equivalent.} but SLOCC equivalent to $|GHZ\ran$ state (the SLOCC operator may be chosen as $\otimes^3_1A_k$ with $A_k=(-1/\sqrt[6]{3})[|0\ran(\lan0|+\lan1|)+|1\ran(\omega\lan0|+\omega^2\lan1|)]$, $\omega$ being a complex cubic root of unity).

Interestingly, however, the four qubit $G$ state is LU equivalent to the $GHZ$ state\footnote{Writing in $|\pm\ran$ basis, i.e., the LU transformation $|0\ran\to|+\ran$ and $|1\ran\to|-\ran$ yield $|G_4\ran=1/\sqrt{2}(|+\ran^{\otimes4}-|-\ran^{\otimes4})$, where $|\pm\ran=(1/\sqrt{2})(|0\ran\pm|1\ran)$ \cite{ausen031}.} and hence cannot be determined by its RDMs!

 For $N\ge5$, relaxing the normalization, we can write \bq\label{gnaswwbar1}|G_N\ran=|00\ran|W_{N-2}\ran+|\psi\ran|GHZ_{N-2}\ran+|11\ran|\bar{W}_{N-2}\ran\eq where $|\psi\ran=|01\ran+|10\ran$. Thus any (normalized) bipartite RDM of $|G_N\ran$ is given by $$\frac{1}{2N}\left(
                                                                                                 \begin{array}{cccc}
                                                                                                   N-2 & 0 & 0 & 0 \\
                                                                                                   0 & 2 & 2 & 0 \\
                                                                                                   0 & 2 & 2 & 0 \\
                                                                                                   0 & 0 & 0 & N-2 \\
                                                                                                 \end{array}
                                                                                               \right)$$
 This matrix has three non-zero eigenvalues whereas any bipartite RDM of $GGHZ$ (or its LU equivalent) has only two non-zero eigenvalues. Since, unitary transformation can not change the eigenvalues, $|G_N\ran$ is not LU equivalent to $|GGHZ\ran$. Therefore, from Walck and Lyons' result \cite{wl2009} it follows that $|G_N\ran$ is uniquely determined by its (all) $(N-1)$-partite RDMs. It is amazing that six years before the general result of \cite{wl2009}, the authors of \cite{ausen031} had proven explicitly and argued that ``$|G_N\ran$ does not belongs to the $|GHZ\ran$ family".

 For $N\ge6$, we have the following stronger result:

\begin{theorem}\label{th4}For $N\ge6$, the $N$-qubit generic $G$ state \bq\label{geng}|GG_N\ran=\sum_{K=1}^N (a_K|01_K0\ldots0\ran+ b_K|10_K1\ldots1\ran) \eq (with $\sum (|a_K|^2+|b_K|^2)=1$, $a_Kb_K\ne0$)  is uniquely determined, among arbitrary states, by its $(N-2)$-partite RDMs, but can not be determined by lower order RDMs.\end{theorem}

\textbf{Proof:} Following the proof of Theorem~\ref{th1}, it can be shown that $|GG_6\ran$ is determined by only three RDMs $\rho^{1234},\rho^{1235},\rho^{1236}$. However, for $N\ge7$, there is a very simple proof which is outlined below:

Since $|GG_N\ran$ has no basis term containing two  or three 1 (and rest zeros), comparing the coefficients of $|01_K0\ldots0\ran\lan01_K0\ldots0|$ in the RDMs, it follows that $r_{(01_K0\ldots0)(01_K0\ldots0)}=|a_K|^2$. Similarly, (interchanging 0 and 1) $r_{(11\ldots10_K1\ldots1)(1\ldots10_K1\ldots1)}=|b_K|^2$. Then by normalization, it follows that the mixed $\rho$ should have the same form as $|GG_N\ran\lan GG_N|$ and share the same diagonals. Finally the off diagonals: it follows trivially (e.g., by comparing $|01_J0\ldots0\ran\lan01_K0\ldots0|$ etc.) $r_{(01_J0\ldots0)(01_K0\ldots0)}=a_J\bar{a}_K, r_{(10_J1\ldots1)(10_K1\ldots1)}=b_J\bar{b}_K$ and $r_{(01_J0\ldots0)(10_K1\ldots1)}=a_J\bar{b}_K$.  The only remaining off diagonals $r_{(01_J0\ldots0)(10_J1\ldots1)}$ are found to be $a_J\bar{b}_J$ by considering PSD of the principal minor consisting of rows and columns $(01_J0\ldots0), (01_K0\ldots0)$ and $(10_J1\ldots1)$.

To prove that $|GG_N\ran$ can not be determined by lower order RDMs, it is sufficient to note that it shares all $(N-3)$-partite RDMs with the following two states \bqa |G'_N\ran&=&|W\ran-|\bar{W}\ran\nonumber\\
\text{and}\quad\rho&=&|W\ran\lan W|+|\bar{W}\ran\lan\bar{W}|,\nonumber\eqa
where the two unnormalized states are given by  $|W\ran=\sum_{K=1}^N a_K|01_K0\ldots0\ran$ and $|\bar{W}\ran=\sum_{K=1}^N b_K|10_K1\ldots1\ran$.~~\hfill $\blacksquare$

The next obvious question would be the optimal number of RDMs required.  Well, it can be proved that for $N\ge6$, only two RDMs (e.g., $\rho^{123\ldots (N-3)(N-2)}, \rho^{345\ldots (N-1)N}$) are required and this is the optimal number.  Here, in the last step (the PSD completion step), instead of using so many third-ordered principle minors, we may consider the fourth-ordered one consisting of the following rows and columns $(01_J0\ldots0),(100\ldots0),(101\ldots1),(10_k1\ldots1)$ and we have to use the result that the following matrix is PSD iff $a=b=c=1$\footnote{First note that the principal minor $A[134]=-|a-b|^2$. So, for PSD, $a=b$. Similarly from $A[124]$, $b=c$. Now, $A[123]=-|a-1|^2$. So for PSD, we must have $a=1$.}:$$\left[
                                                                     \begin{array}{cccc}
                                                                       1 & 1 & a & b \\
                                                                       1 & 1 & 1 & c \\
                                                                       \bar{a} & 1 & 1 & 1 \\
                                                                       \bar{b} & \bar{c} & 1 & 1 \\
                                                                     \end{array}
                                                                   \right].$$

\section{Discussion and Conclusion}\label{end}
First of all we want to mention that though we have studied \emph{generic} classes of states, except for the $W$ class ($|GW\ran$), the term `generic' means that the coefficients are arbitrary complex numbers and nothing else, whereas for the $W$-class, it includes all `$W$-type' states i.e., all states which are SLOCC equivalent to the $W$ state. We note that under SLOCC, the generic Dicke state (\ref{defdicke}) transforms as \bq\label{dickeslocc}|GD_N^{(\ell)}\ran\rightarrow\sum_{k=0}^{\ell} |GD_N^{(k)}\ran.\eq Thus, under SLOCC, $|GD_N^\ell\ran$ does not preserve its \emph{minimal} form. So, it is almost impossible to check their minimal reducibility by the present technique. However, we should emphasize that this is not a shortcoming of the technique. Under SLOCC, most states change drastically and for $N\ge4$, there are uncountable number of SLOCC inequivalent states \cite{gourwallachjmp10}. So, it is unusual to expect to explicitly express each member and then characterize the classes case by case. That is why we have considered the generic classes like this. It is also worth pointing out that each such class is also composed of several (uncountable number of) SLOCC inequivalent states. For example, using the criterion of \cite{gourwallachjmp10}, it follows immediately that the two members from the family $|GD_4^2\ran$ \bqa |\psi\ran&=&a|3\ran+b|5\ran+c|6\ran+d|9\ran+e|10\ran+f|12\ran\nonumber\\\text{and }|\phi\ran&=&a'|3\ran+b'|5\ran+c'|6\ran+d'|9\ran+e'|10\ran+f'|12\ran\nonumber\eqa are SLOCC inequivalent unless $af(cd-be)=a'f'(c'd'-b'e')$. This also implies that all members of the family $|GD_4^2\ran$ with $af(cd-be)\ne0$ are SLOCC inequivalent to $|D_4^2\ran$.

Because of the powerful result of \cite{wl2009}, it is now very easy to check whether any $N$-qubit pure state has reducible correlations or not--we just need to check whether the given state is LU equivalent to $|GGHZ\ran$ and this question of LU equivalence has been recently solved in \cite{bkrausprl10}. However, our aim is not just to determine the reducibility, but following the spirit of the original work \cite{lpw2002, lw2002}, to determine how far we can reduce the correlations. This question has similar notion with separability and $K$-separability. To resolve it, we have to characterize all classes of states that can be determined by $K$-partite RDMs. Though some constraints can be derived easily, we are not yet able to address this issue. Very recently, the authors of \cite{ushaothers} have applied Majorana representation to determine the reducibility of symmetric classes of states. This approach may give some insight to solve the problem.

As mentioned earlier, recently the correlation structure of $|G_N\ran$ has attracted much attention. The authors of \cite{ausen031} had previously shown that the ``higher order correlation is reducible to lower order ones" and thus \emph{$|G_N\ran$ is weakly correlated} than $|GHZ\ran$. It follows from our result that we can lower one more level thereby making the correlation even weaker, but many much stronger than that of $W$-type state itself. It indeed is surprising that the correlation information in $|GG_N\ran$ is \emph{imprinted} in just two $(N-2)$-partite RDMs. A byproduct of this result is that all such states are different (LU inequivalent) from $|GGHZ\ran$.

 To conclude, we have shown that all multiqubit pure states which are SLOCC equivalent to the $N$-qubit $W$ state, are uniquely determined by their bipartite marginals. So, from the parts and whole perspective, we can say that these states contain information essentially at the bipartite level. Moreover, only ($N$-1) number of bipartite RDMs having one party common to all, are required and this number is optimal. Entanglement (by construction of any measure) is always preserved under LU but in general not under SLOCC. The same holds for (ir)reducibility in the case of $GHZ$ (and so for generic) states. However, on the contrary, for the $W$ states, it is rather surprising that reducibility is preserved under SLOCC.  Prior to this work, even the reducibility of all LU equivalent $W$ states was not known. We hope our results will help to understand and explore further the correlation structure of W-type states.

\appendix*
\section{Proof of Theorem~\ref{th2}}
 As usual, let an $N$-qubit (generically mixed) state $\rho_M$, as given by Eq.~(\ref{grp1}) be such that $\rho^{K(K+1)}_M=\rho^{K(K+1)}_{GW}\quad\forall K=1(1)(N-1)$. We will show that $\rho_M=|GW\ran\lan GW|$.

 First we note that no basis of ${\rho}_M$ can have two consecutive 1 (means $r_{(i_1i_2\ldots i_{K-1}1_K1_{K+1}i_{K+2}\ldots i_N)(j_1j_2\ldots j_N)}=0$). Next we will show that no basis of $\rho_M$ can have the sequence 101 i.e. $r_{(i_1\ldots 1_K0_{K+1}1_{K+2}\ldots i_N)(j_1j_2\ldots j_N)}=0$. For simplicity, let us first take $K=1$ and the other cases will follow similarly. So, comparing the diagonal elements $|01\ran\lan01|$, $|10\ran\lan10|$ and the off-diagonal elements $|01\ran\lan10|$ from $\rho^{12}_M$ and $\rho^{12}_{GW}$ we have (keeping in mind that no basis can have two consecutive 1)

\begin{subequations}
\label{allequations}
\bqa \sum_{i_4,i_5,\ldots,i_N=0}^1r_{(010i_4i_5\ldots i_N)(010i_4i_5\ldots i_N)} &=&|c_2|^2\label{diag1}\\
\sum_{i_3,i_4,\ldots,i_N=0}^1 r_{(10i_3i_4\ldots i_N)(10i_3i_4\ldots i_N)}&=&|c_1|^2\label{diag2}\\
\sum_{i_4,i_5,\ldots,i_N=0}^1 r_{(010i_4i_5\ldots i_N)(100i_4i_5\ldots i_N)}&=&c_2\bar{c}_1\label{diag3}
\eqa
\end{subequations}

Considering absolute values in (\ref{diag3}), we have \bq\sum_{i_4,i_5,\ldots,i_N=0}^1 |r_{(010i_4i_5\ldots i_N)(100i_4i_5\ldots i_N)}|\ge|c_2||c_1|\label{lowbounddiag3}\eq

 By the PSD property (iii),
\begin{widetext}
$|r_{(010i_4i_5\ldots i_N)(100i_4i_5\ldots i_N)}|\le \sqrt{r_{(010i_4i_5\ldots i_N)(010i_4i_5\ldots i_N)}r_{(100i_4i_5\ldots i_N)(100i_4i_5\ldots i_N)}}$. So summing up (each sum varies over $i_4,i_5,\ldots,i_N=0$ to 1),
\begin{subequations}
\label{allequations}
\bqa \sum |r_{(010i_4\ldots i_N)(100i_4\ldots i_N)}|&\le&\sum \sqrt{r_{(010i_4\ldots i_N)(010i_4\ldots i_N)}r_{(100i_4\ldots i_N)(100i_4\ldots i_N)}}\label{diag21}\\
&\le&\sqrt{(\sum r_{(010i_4\ldots i_N)(010i_4\ldots i_N)})(\sum r_{(100i_4\ldots i_N)(100i_4\ldots i_N)})}\label{diag22}\\
&\le&|c_2||c_1|\label{diag23}\eqa
\end{subequations}
\end{widetext}
where in (\ref{diag22}) we have used Schwartz inequality and in (\ref{diag23}), we have used (\ref{diag1}) and (\ref{diag2}).  It follows from (\ref{lowbounddiag3}) and (\ref{diag23}) that all inequalities in (\ref{lowbounddiag3}) and (A.3) should be equalities.  For equality in (\ref{diag23}) we must have \bq\label{i3zero}\sum_{i_4,\ldots,i_N=0}^1 r_{(100i_4\ldots i_N)(100i_4\ldots i_N)}=|c_1|^2\eq which together with (\ref{diag2}) implies that $r_{(101i_4i_5\ldots i_N)(101i_4i_5\ldots i_N)}=0$. Comparing $\rho^{K(K+1)}_M$ and $\rho^{K(K+1)}_{GW}$, in a similar way we can prove that no basis of $\rho_M$ can contain the sequence $101$ i.e., \bq\label{nobasis101} r_{(i_1i_2\ldots i_{K-1}1_{K}0_{K+1}1_{K+2}i_{K+3}\ldots i_N)(j_1j_2\ldots j_N)}=0\eq Thus it follows immediately that (\ref{diag3}) would reduce to \bq\label{i4zero}\sum_{i_5,i_6,\ldots,i_N=0}^1 r_{(0100i_5\ldots i_N)(1000i_5\ldots i_N)}=c_2\bar{c}_1\eq
as well as $r_{(1001i_5\ldots i_N)(j_1j_2\ldots j_N)}=0$. Similarly, it follows that $i_5=0$, $i_6=0$  by considering $\rho^{23}$, $\rho^{34}$ and so on. For an illustration, the iterations go like: \bqa \rho^{23}:~\sum_{i_6,\ldots,i_N=0}^1 r_{(00100i_6\ldots i_N)(01000i_6\ldots i_N)}&=&c_3\bar{c}_2\nonumber\\
\rho^{34}:~\sum_{i_7,\ldots,i_N=0}^1 r_{(000100i_7\ldots i_N)(001000i_7\ldots i_N)}&=&c_4\bar{c}_3\nonumber\eqa and so on. The iteration stops at $\rho^{(N-1)N}$ and we will have\bqa r_{(001_K00\ldots0)(001_K00\ldots0)}&=&|c_K|^2\nonumber\\r_{(00_K1_{K+1}00\ldots0)(01_K0_{K+1}0\ldots0)}&=&c_{K+1}\bar{c}_K\eqa Then by normalization, $r_{(00\ldots0)(00\ldots0)}=|c_0|^2$. Finally, the off diagonal elements $r_{(001_J0\ldots0)(01_K0\ldots0)}$ are found to be $c_J\bar{c}_K$ by the repeated applications of the third-ordered PSD completion.\hfill $\blacksquare$


\begin{thebibliography}{50}

\bibitem{dvc2000} W. D\"{u}r, G. Vidal and J. I. Cirac, \href{http://dx.doi.org/10.1103/PhysRevA.62.062314}{Phys Rev. A {\bf62}, 062314 (2000)}.

\bibitem{lpw2002} N. Linden, S. Popescu and W. K. Wootters, \href{http://dx.doi.org/10.1103/PhysRevLett.89.207901}{Phys. Rev. Lett. {\bf89}, 207901 (2002)}.

\bibitem{lw2002} N. Linden and W. K. Wootters, \href{http://dx.doi.org/10.1103/PhysRevLett.89.277906}{Phys. Rev. Lett. 89,  277906 (2002)};
A. Higuchi, A. Sudbery and J. Szulc, \href{http://dx.doi.org/10.1103/PhysRevLett.90.107902}{Phys. Rev. Lett. {\bf90}, 107902 (2003)};
L. Di\'{o}si, \href{http://dx.doi.org/10.1103/PhysRevA.70.010302}{Phys. Rev. A {\bf70}, 010302 (2004)};
Y. Feng, R. Duan and M. Ying, \href{http://www.rintonpress.com/journals/qiconline.html#v9n1112}{Quant. Inf. Comp. {\bf9}, 0997 (2009)}.

\bibitem{jl2005} N. S. Jones  and N. Linden, \href{http://dx.doi.org/10.1103/PhysRevA.71.012324}{Phys. Rev. A \textbf{71} 012324 (2005)}.

\bibitem{wl2009} S. N. Walck and D. W. Lyons, \href{http://dx.doi.org/10.1103/PhysRevLett.100.050501}{Phys. Rev. Lett. \textbf{100}, 050501 (2008)};
\href{http://dx.doi.org/10.1103/PhysRevA.79.032326}{Phys. Rev. A \textbf{79}, 032326 (2009)}.

\bibitem{z2008} D. L. Zhou, \href{http://dx.doi.org/10.1103/PhysRevLett.101.180505}{Phys. Rev. Lett. {\bf101}, 180505 (2008)};
\href{http://dx.doi.org/10.1103/PhysRevA.80.022113}{Phys. Rev. A {\bf80}, 022113 (2009)}.

\bibitem{yangeisert09} D. Yang and J. Eisert, \href{http://dx.doi.org/10.1103/PhysRevLett.103.220501}{Phys. Rev. Lett. {\bf103}, 220501 (2009).}

\bibitem{pr2009} P. Parashar and S. Rana, \href{http://dx.doi.org/10.1103/PhysRevA.80.012319}{Phys. Rev. A \textbf{80}, 012319 (2009)}.

\bibitem{prftc2009} P. Parashar and S. Rana, \href{http://dx.doi.org/10.1088/1751-8113/42/46/462003}{J. Phys. A: Math. Theor. \textbf{42}, 462003  (2009)}.

\bibitem{nielsen99} M. A. Nielsen, \href{http://dx.doi.org/10.1103/PhysRevLett.83.436}{Phys. Rev. Lett. {\bf83}, 436 (1999).}

\bibitem{ut10} K. Uyanik and S. Turgut, \href{http://dx.doi.org/10.1103/PhysRevA.81.032306}{Phys. Rev. A {\bf81}  032306 (2010)}.

\bibitem{winter10} L. Chen, E. Chitambar, R. Duan, Z. Ji and A. Winter, \href{http://dx.doi.org/10.1103/PhysRevLett.105.200501}{Phys. Rev. Lett. {\bf105},
200501 (2010);} L. Chen and M. Hayashi, \href{http://dx.doi.org/10.1103/PhysRevA.83.022331}{Phys. Rev. A {\bf83}, 022331 (2011).}

\bibitem{exp} P. Agrawal and A. Pati, \href{http://dx.doi.org/10.1103/PhysRevA.74.062320}{Phys. Rev. A {\bf74},  062320 (2006)};\\
L. Li and D. Qiu, \href{http://dx.doi.org/10.1088/1751-8113/40/35/010}{J. Phys. A: Math. Theor. {\bf40} 10871 (2007)};
Z. H. Peng, J. Zou and X. J. Liu, \href{http://dx.doi.org/10.1140/epjd/e2010-00106-8}{Eur. Phys. J. D {\bf58}, 403 (2010)}.

\bibitem{kt2010} S. Kintas and S. Turgut, \href{http://dx.doi.org/10.1063/1.3481573}{J. Math. Phys. {\bf51}, 092202 (2010)}.

\bibitem{hj1985} See e.g., R. A. Horn and C. R. Johnson, \emph{Matrix analysis}, published by Cambridge University Press, USA, 1985.

\bibitem{dliepl09} D. Li, X. Li, H. Huang and X. Li, \href{http://dx.doi.org/10.1209/0295-5075/87/20006}{EPL {\bf87}, 20006 (2009).}

\bibitem{ausen03} A. Sen(De), U. Sen, M. Wie\'{s}niak, D. Kaszlikowski and M. \.{Z}ukowski, \href{http://dx.doi.org/10.1103/PhysRevA.68.062306}{Phys. Rev. A {\bf68}, 062306 (2003);} D. Kaszlikowski, A. Sen(De), U. Sen, V. Vedral and A. Winter, \href{http://dx.doi.org/10.1103/PhysRevLett.101.070502}{Phys. Rev. Lett. {\bf101}, 070502 (2008).}

\bibitem{ausen031} A. Sen(De), U. Sen and M. \.{Z}ukowski, \href{http://dx.doi.org/10.1103/PhysRevA.68.032309}{Phys. Rev. A {\bf68}, 032309 (2003).}

\bibitem{bennett11} C. H. Bennett, A. Grudka, M. Horodecki, P. Horodecki and R. Horodecki, \href{http://dx.doi.org/10.1103/PhysRevA.83.012312}{Phys. Rev. A {\bf83}, 012312 (2011).}

\bibitem{dsrar} A. R. Usha Devi, Sudha and A. K. Rajagopal, \emph{Interconvertibility and irreducibility of permutation symmetric three qubit pure states},
\href{http://arxiv.org/abs/1002.2820}{arXive:1002.2820} [quant-ph].

\bibitem{gourwallachjmp10} G. Gour and N. R. Wallach, \href{http://dx.doi.org/10.1063/1.3511477}{J. Math. Phys. {\bf51}, 112201 (2010).}

\bibitem{bkrausprl10} B. Kraus, \href{http://dx.doi.org/10.1103/PhysRevLett.104.020504}{Phys. Rev. Lett. {\bf104}, 020504 (2010).}

\bibitem{ushaothers} A. R. Usha Devi, Sudha and A. K. Rajagopal, \href{http://arxiv.org/abs/1003.2450}{arXiv:1003.2450;} Quantum Inf Process, \href{http://dx.doi.org/10.1007/s11128-011-0280-8}{DOI 10.1007/s11128-011-0280-8.}

\end{thebibliography}
\end{document}